\documentstyle[twocolumn,floats,epsfig,prl,aps]{revtex}

\tightenlines

\begin{document}

\draft
\title{Droplet Nucleation and Domain Wall Motion in a Bounded Interval}
\author{Robert S. Maier${}^{1,2}$ and D.~L. Stein${}^{2,1}$}
\address{${}^{1}$Mathematics and ${}^{2}$Physics Departments, University of
Arizona, Tucson, Arizona 85721}

\maketitle

\begin{abstract}
We study a spatially extended model of noise-induced magnetization
reversal: a~classical Ginzburg--Landau theory, restricted to a bounded
interval and perturbed by weak spatiotemporal noise.  We~compute the
activation barrier and Kramers prefactor.  As~the interval length
increases, a~transition between activation regimes occurs, at~which the
prefactor diverges.  We~relate this to transitions that occur in
low-temperature quantum field theory.
\end{abstract}


\narrowtext

The effect of noise on spatially extended classical-mechanical systems has
become a subject of intense investigation~\cite{GarciaOjalvo99}.  Noise, by
which is meant local fluctuations of thermal or other origin, may be
spatiotemporal: it~may vary randomly in space as~well as time.

A typical problem is the determination of the extent to which
spatiotemporal noise can induce transitions between the stable states of a
system described by a nonlinear field equation, or induce such a system to
escape from a metastable state.  Langer~\cite{Langer67} and B\"uttiker and
Landauer~\cite{Buttiker79and81} studied one-dimensional versions of
this problem, in the limit of an infinite-size spatial region.  Recently,
finite size effects have become a subject of study and simulation.
Micromagnetics is one application area.  The thermally activated
magnetization reversal of nanoscale magnets is described by coupled
Landau--Lifschitz--Gilbert equations perturbed by spatiotemporal
noise~\cite{Brown00}.  An~interesting question is how the effects of
spatial extent cause magnetization reversal to differ from the
better-understood `zero-dimensional' case of a single-domain particle, with
the noise taken to have no spatial dependence~\cite{Boerner98}.  Another
area where activation by spatiotemporal noise is important is the formation
of spatially localized structures in
electroconvection~\cite{Bisang98andTu97}.

In this Letter, we quantify the effects of weak spatiotemporal noise on an
overdamped bistable quartic (i.e., double well) classical Ginzburg--Landau
field theory, in a one-dimensional region: the bounded interval~$[0,L]$.
This model is similar to that of Langer~\cite{Langer67}, and has much in
common with the sine-Gordon model of B\"uttiker and
Landauer~\cite{Buttiker79and81}.  For sufficiently large~$L$, the model has
two stable states, which are states of positive and negative magnetization.
In~the Kramers weak-noise limit, in which noise-activated magnetization
reversals become exponentially rare, we compute the reversal rate~$\Gamma$.
We also determine the `optimal trajectories', in the model's
infinite-dimensional state space, for magnetization reversal.  They are
greatly affected by the choice of boundary conditions.  If~periodic
boundary conditions are used, it is most likely that reversal will proceed
via the nucleation of a droplet within a pair of Bloch walls; but for
Dirichlet or Neumann boundary conditions, a single wall will form at $x=0$
or~$x=L$, and sweep across the interval.  Faris and Jona-Lasinio worked~out
a mathematically rigorous `large deviation theory' of the Dirichlet-case
reversal~\cite{Faris82andMartinelli89}.

Reversals fall~off according to $\Gamma\sim \Gamma_0\exp(-\Delta
E/\epsilon)$, where $\epsilon$~is the noise strength, $\Delta E$ is the
activation barrier between the two states, and $\Gamma_0$~is the Kramers
prefactor.  We~show how to compute the prefactor in closed form, as a
function of~$L$, for all three boundary conditions, and in the Dirichlet
case, work it out in~full.  Langer computed it in the $L\to\infty$ limit of
the periodic case, so this is a significant advance.  The prefactor
contains a quotient of infinite-dimensional fluctuation determinants.
Since the work of Coleman~\cite{Coleman79}, the quantum field theory
community has known how to compute such quotients as $L\to\infty$.  With
the regularization technique of McKane and Tarlie~\cite{McKane95}, we can
handle the $L<\infty$ case, as~well.

A feature of our prefactor computation is the calculation of the unstable
eigenvalue of the model's deterministic dynamics, at the transition state
between its two stable states.  The Kramers prefactor differs significantly
from the analogous low-temperature quantum-mechanical tunneling prefactor,
in that it involves this
eigenvalue~\cite{Affleck81andWolynes81andGrabert84}.  Calculating it in the
Dirichlet case is difficult.  We~express it in terms of a mid-band
eigenvalue of the $l=2$ Lam\'e Hamiltonian \cite{Harrington78,Li00}.  Using
a dispersion relation developed for this
Hamiltonian~\cite{Maierinpreparation}, we compute the eigenvalue, and hence
the Dirichlet-case prefactor.

Our most striking discovery has to do with the phase structure of the
weak-noise limit.  In~a sense that can be made precise, the $\epsilon\to0$
limit of the quartic Ginzburg--Landau theory has a second-order phase
transition at $L=2\pi$ (Dirichlet, periodic cases) or $L=\pi$ (Neumann
case), in dimensionless units.  At~these critical values of the interval
length, the Kramers prefactor {\em diverges\/}.  This is due~to the
bifurcation of the transition state, as $L$~increases through criticality.
A~zero-field `sphaleron' configuration, which serves as transition state
when $L$~is small, bifurcates into a degenerate pair of `periodic
instantons'.  Similar bifurcations have been studied by the quantum field
theory community~\cite{Chudnovsky92,Kuznetsov97,Frost99,Espichan00}.  Until
now, the importance of these bifurcations to classical activation by
spatiotemporal noise has not been fully realized.

{\em Model and Phenomenology.}---In specifying the overdamped
Ginzburg--Landau model, we follow Faris and
Jona-Lasinio~\cite{Faris82andMartinelli89}.  On~$[0,L]$, a classical field
$\phi=\phi(x,t)$ is evolved by the stochastic Ginzburg--Landau equation
\begin{equation}
\label{eq:Langevin}
\dot\phi = M\phi'' + \mu \phi - \lambda \phi^3 + \epsilon^{1/2}\xi(x,t),
\end{equation}
where $\xi(x,t)$ is unit-strength spatiotemporal white noise, satisfying
$\langle\xi(x_1,t_1)\xi(x_2,t_2)\rangle=\delta(x_1-x_2)\delta(t_1-t_2)$.
We~set $M=\mu=\lambda=1$, or equivalently use dimensionless units, in which
the length unit is the nominal coherence length $\sqrt{M/\mu}$, the field
strength unit is $\sqrt{\mu/\lambda}$, etc.

In the absence of noise, the time-independent solutions
of~(\ref{eq:Langevin}) include the sphaleron $\phi\equiv0$, and
$\phi\equiv\pm1$; the latter are the magnetized states in the Neumann and
periodic cases.  In~the Dirichlet case, $\phi(x=0)=\phi(x=L)=0$ is imposed,
so the magnetized states necessarily differ.  It~is easy to check that if
$\epsilon=0$, the so-called periodic instanton $\phi=\phi_{{\rm
inst},m}(x)$ is a time-independent solution of~(\ref{eq:Langevin}) for
any~$m$ in the range $0<m<1$.  Here (cf.~\cite{Espichan00})
\begin{equation}
\phi_{{\rm inst},m}(x) \equiv 
\sqrt{\frac{2m}{m+1}}\,{\rm sn}(x/\sqrt{m+1} \mid m),
\end{equation}
where ${\rm sn}(\bullet\mid m)$ is the Jacobi elliptic function with
parameter~$m$, which has quarter-period ${\bf K}(m)$~\cite{Abramowitz65}.
This quarter-period decreases to~$\pi/2$ as~$m\to0^+$, in which limit ${\rm
sn}(\bullet\mid m)$ degenerates to $\sin(\bullet)$, and it increases to
infinity as~$m\to1^-$.  One would expect the Dirichlet-case stable
magnetized states to be $\pm\phi_{{\rm inst},m_{s,D}}$, with
$m_{s,D}$~determined implicitly by the condition $\phi(x=L)=0$, i.e., by
the half-period condition $2\sqrt{m_{s,D}+1}\,{\bf K}(m_{s,D})=L$.
However, this is so only if $L>\pi$.  If $L\le\pi$, there is no solution
for~$m_{s,D}$ in the range $0<m_{s,D}<1$, and the Dirichlet-case model is
monostable with the zero-field configuration as its stable state, rather
than bistable.  Bistability disappears when $L\to\pi^+$ and
$m_{s,D}\to0^+$.

Noise-activated magnetization reversal, for weak noise, proceeds with high
likelihood along an optimal trajectory in the model's infinite-dimensional
state space that goes `uphill' from a magnetized state to an intermediate
transition state.  In~a zero-dimensional approximation, this is the
sphaleron.  However, on physical grounds one expects a transition state to
be an unstable field configuration with two droplets, i.e., with half the
interval occupied by each magnetization value.  Mathematically, this can
arise as follows.  In~the Dirichlet case, the sphaleron undergoes a
pitchfork bifurcation into the degenerate magnetized states $\pm\phi_{{\rm
inst},m_{s,D}}$ when $L$~is increased through~$\pi$.  When $L$~is increased
through~$2\pi$, it bifurcates again, into a degenerate pair of unstable
states $\pm\phi_{{\rm inst},m_{u,D}}$.  Unlike the magnetized states, these
transition states are positive on half the interval and negative on the
other.  $m_{u,D}$~is determined from~$L$ by the condition $\phi(x=L)=0$,
i.e., by the quarter-period condition $4\sqrt{m_{u,D}+1}\,{\bf
K}(m_{u,D})=L$.  

The same bifurcation occurs if periodic boundary conditions are used, but
the new transition state is infinitely degenerate: it~is $\phi_{{\rm
inst},m_{u,D}}$, translated arbitrarily.  It~is easy to see that in the
Neumann case, the bifurcation occurs at $L=\pi$ rather than $L=2\pi$, with
the Neumann parameter $m_{u,N}$ determined by the half-period condition
$2\sqrt{m_{u,N}+1}\,{\bf K}(m_{u,N})=L$.  In~fact, $m_{u,N}=m_{s,D}$.
To~satisfy Neumann boundary conditions, the field configurations
$\pm\phi_{{\rm inst},m_{u,N}}$ must be shifted by~$L/2$.

\begin{figure}[t]
\centerline{\epsfig{file=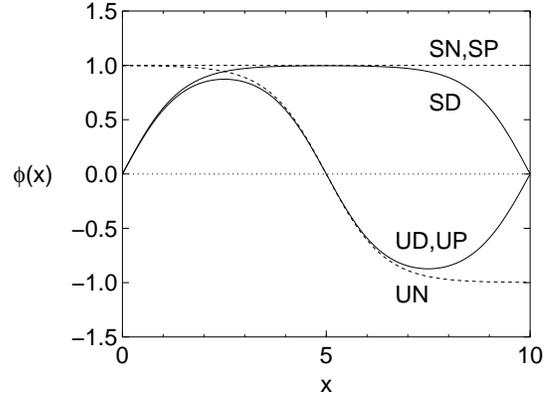,width=2.8964in}}
\vskip0.15in
\caption{Stable magnetized states~(S) and transition states~(U), for
Dirichlet~(D), Neumann~(N), and periodic~(P) boundary conditions, if
$L=10$.  Each state may be multiplied by~$-1$, and UP~may be shifted
arbitrarily.}
\end{figure}

Figure~1 displays the magnetization and transition states for each boundary
condition.  As~$L\to\infty$, the Bloch walls between magnetization values
increasingly acquire the standard hyperbolic tangent form, since ${\rm
sn}(\bullet\mid m)$ degenerates to $\tanh(\bullet)$ as~$m\to1^-$.

The zero-noise dynamics of this model are of the gradient form
$\dot\phi=-\delta{\cal H}/\delta\phi$, with the energy functional
\begin{equation}
\label{eq:energy}
{\cal H}[\phi]\equiv
\int_0^L \left[(\phi')^2 /2 + \phi^4/4 - \phi^2/2 + 1/4 \right]\,dx.
\end{equation}
The energy of each magnetization and transition state is readily computed
from~(\ref{eq:energy}).  The states $\phi\equiv\pm1$ have zero energy, and
the sphaleron has energy~$L/4$.  The energy of each periodic instanton state
$\pm\phi_{{\rm inst},m}$ turns~out to~be
\begin{equation}
\frac{L}{12}\left[
\frac{8}{(m+1)}\,
\frac{{\bf E}(m)}{{\bf K}(m)} - \frac{(1-m)(3m+5)}{(m+1)^2}
\right],
\end{equation}
where ${\bf E}(m)$ is the second complete elliptic
integral~\cite{Abramowitz65}.  

Figure~2 plots the activation barrier~$\Delta E$ as a function of~$L$, for
each boundary condition.  The second derivative of each $\Delta E$ function
is discontinuous at the value of~$L$ at which bifurcation occurs.
As~$L\to\infty$, the value of~$\Delta E$ converges to the energy of a Bloch
wall (Dirichlet, Neumann cases), or two Bloch walls (periodic case).
It~follows from~(\ref{eq:energy}) that the energy of a Bloch wall, of the
$m\to1^-$ limiting form $\tanh (x/\sqrt{2})$, is $2\sqrt{2}/3\approx0.943$.

\begin{figure}[t]
\centerline{\epsfig{file=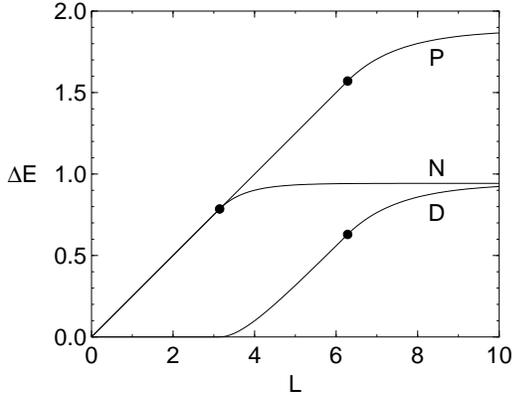,width=2.7606in}}
\vskip0.15in
\caption{The activation barrier~$\Delta E$, for the cases of Dirichlet~(D),
Neumann~(N), and periodic~(P) boundary conditions.  Bullets indicate
criticality ($L=2\pi$ for D and~P; $L=\pi$ for~N).}
\end{figure}

{\em Determinant Quotients.}---The formula for the Kramers
prefactor~$\Gamma_0$ of an overdamped multidimensional system driven by
weak white noise is well known.  Suppose the system has a stable
state~${\bbox\varphi}_s$ and a transition state~${\bbox\varphi}_u$, with a
single unstable direction.  Let ${\bf\Lambda}_s$ and~${\bf\Lambda}_u$
denote the system's linearized noiseless dynamics at ${\bbox\varphi}_s$
and~${\bbox\varphi}_u$, so that to leading order, the state
${\bbox\varphi}={\bbox\varphi}_s+{\bbox\eta}$ evolves by
$\dot{\bbox\eta}=-{\bf\Lambda}_s{\bbox\eta}$, and
${\bbox\varphi}={\bbox\varphi}_u+{\bbox\eta}$ by
$\dot{\bbox\eta}=-{\bf\Lambda}_u{\bbox\eta}$.  Then~\cite{McClintock89a}
\begin{equation}
\label{eq:prefactor}
\Gamma_0 = 
\frac1{2\pi}\sqrt{\left|\Upsilon\right|}\,\,\left|\lambda_{u,1}\right|
\equiv
\frac1{2\pi}
\sqrt{\left|\frac{\det{\bf\Lambda}_s}{\det{\bf\Lambda}_u}\right|}
\,\,\left|\lambda_{u,1}\right|,
\end{equation}
where $\lambda_{u,1}$ is the only negative eigenvalue of~${\bf\Lambda}_u$.
The corresponding eigenvector~${\bbox\eta}_{u,1}$ is the direction along
which the optimal trajectory approaches the transition state.
If~$\lambda_{s,1}$ denotes the smallest eigenvalue of~${\bf\Lambda}_s$, the
corresponding eigenvector~${\bbox\eta}_{s,1}$ will be the direction along
which the trajectory extends from the stable state.

Linearizing the noiseless version of~(\ref{eq:Langevin}) at a stationary
state~$\phi_0$ (either a stable or a transition state) yields
\begin{equation}
\dot\eta = -\hat\Lambda[\phi_0]\,\eta \equiv
-\left[-{d^2}/{dx^2} + (-1 + 3\phi_0^2)\right]\eta.
\end{equation}
So $\Gamma_0$ depends on the spectrum of the $\hat\Lambda$~operators
associated with the stable and transition states.  In the Dirichlet case,
the formal determinant quotient~$\Upsilon$ can be computed by the now
standard technique of Coleman~\cite{Coleman79,McKane95}.  If $L>2\pi$, let
$\eta_{s,*}$ and~$\eta_{u,*}$ be the solutions on~$[0,L]$ of the
homogeneous equations $\hat\Lambda[\phi_{{\rm inst},m_{s,D}}]\eta=0$ and
$\hat\Lambda[\phi_{{\rm inst},m_{u,D}}]\eta=0$ which satisfy the boundary
conditions $\eta(0)=0$ and $\eta'(0)=1$.  Then, it turns~out,
\begin{equation}
\label{eq:quotient}
\Upsilon_D \equiv \eta_{s,*}(L)/\eta_{u,*}(L).
\end{equation}
is the Dirichlet-case determinant quotient.

Solutions $\eta_{s,*}$ and $\eta_{u,*}$ satisfying these special boundary
conditions may be constructed by a clever trick~\cite{McKane95}:
differentiating the periodic instanton $\phi_{{\rm inst},m}$ with respect
to~$m$, setting $m$ to $m_{s,D}$ and~$m_{u,D}$ respectively, and
normalizing.  This procedure uses the formula for the derivative of ${\rm
sn}(\bullet\mid m)$ with respect to~$m$~\cite{Neville51}.  The result is
\begin{displaymath}
\eta_{c,*}(L) = 
\pm\frac{L}{m_{c,D}^2-m_{c,D}+1}
\left[
\frac{m_{c,D}+1}{1-m_{c,D}}
\,
\frac{{\bf E}(m_{c,D})}{{\bf K}(m_{c,D})}
- 1
\right],
\end{displaymath}
for both $c=s$ and~$c=u$ (`$\pm$' is positive and negative respectively).
Substituting $\eta_{s,*}(L)$ and $\eta_{u,*}(L)$ into~(\ref{eq:quotient})
yields the determinant quotient~$\Upsilon_D$.

The unbifurcated regime, i.e., $\pi<L<2\pi$, must be handled a bit
differently.  Since the transition state is the sphaleron, and not the
periodic instanton $\phi_{{\rm inst},m_{u,D}}$, $\eta_{u,*}(L)$ cannot be
computed from the above formula.  But it is trivial to check that
$\eta_{u,*}(L)$ simply equals $\sin L$.

The Neumann case is treated similarly to the Dirichlet (we omit the
details), but the case of periodic boundary conditions is very different,
at~least in the bifurcated regime $L>2\pi$.  The periodic instanton
transition state is infinitely rather than doubly degenerate, and at any
transition state, the linearized dynamical operator~$\hat\Lambda$ has a
soft `collective mode', with a zero eigenvalue.
Equation~(\ref{eq:prefactor}) must be modified, but the regularization
technique of McKane and Tarlie~\cite{McKane95} can be used to work out the
periodic-case Kramers prefactor~\cite{MaierSteininpreparation}.
It~acquires an $\epsilon^{-1/2}$ factor.  As~a result, the periodic-case
reversal rate becomes non-Arrhenius when $L$~is increased through~$2\pi$.
A~similar non-Arrhenius reversal rate falloff is displayed, in the limit of
zero spatial extent, by the stochastic Landau--Lifschitz--Gilbert
equation~\cite{Boerner98}, due to its space of magnetization values having
a continuous symmetry.

{\em The Unstable Eigenvalue.}---The stumbling block in the analytic
computation of the Dirichlet-case Kramers prefactor is the calculation
of~$\lambda_{u,1}$, the single negative (unstable) eigenvalue of the
deterministic dynamics, linearized at the transition state.  If
$\pi<L<2\pi$ and the transition state is the sphaleron,
$\lambda_{u,1}=\pi^2/L^2-1$ is trivial to verify.  But computing
$\lambda_{u,1}$ and the corresponding eigenfunction $\eta_{u,1}$ when
$L>2\pi$ is much harder.  $\eta_{u,1}$~is of considerable physical
interest, since it describes the way in~which the optimal trajectories
approach the periodic instanton solutions $\pm\phi_{{\rm inst},m_{u,D}}$,
i.e., the way in~which the moving Bloch wall slows to a halt at~$x=L/2$.

Here we sketch the calculation of $\lambda_{u,1}$ and~$\eta_{u,1}$ from the
eigenvalue equation $\hat\Lambda\eta=\lambda\eta$; details will appear
elsewhere~\cite{MaierSteininpreparation}.  Introducing $z\equiv
x/\sqrt{m_{u,D}+1}$, and for simplicity, writing $m$ for~$m_{u,D}$,
converts the equation to
\begin{equation}
\label{eq:Lame}
\left[-d^2/dz^2 + 6m\,{\rm sn}^2(z\mid m)\right]\eta = {\cal E}\eta,
\end{equation}
where the `energy'~$\cal E$ equals $(m+1)(\lambda_{u,1}+1)$.  The interval
$0\le z\le 4{\bf K}(m)$ corresponds to~$[0,L]$.  Equation~(\ref{eq:Lame})
is the $l=2$ Lam\'e equation~\cite{Harrington78,Li00}, which is a
Schr\"odinger equation with a periodic potential, whose lattice constant
is~$2{\bf K}(m)$.  Its Bloch wave spectrum is known to consist of three
energy bands~\cite{Li00}, each extending over the wavenumber range
$-\pi/2{\bf K}(m) \le k \le \pi/2{\bf K}(m)$.

According to Hermite's solution of the Lam\'e equation~\cite{Whittaker27},
Eq.~(\ref{eq:Lame}) has solutions of the form
\begin{equation}
\label{eq:Hermitesoln}
\eta(z)=\prod_{i=1}^2
\left\{
\left[\frac{{\rm H}(z\pm\alpha_i\mid m)}{\Theta(z\mid m)}\right]
    \exp[\mp\, {\rm Z}(\alpha_i\mid m)\,z]
\right\},
\end{equation}
where ${\rm H}$, $\Theta$, and~${\rm Z}$ are the Jacobi eta, theta, and
zeta functions, and $\alpha_1,\alpha_2$ are certain complex constants
determined by~$\cal E$.  Clearly, the wavenumber~$k$ of the
solution~(\ref{eq:Hermitesoln}) equals $\pm{\rm Im}\sum_{i=1}^2 {\rm
Z}(\alpha_i\mid m)$.

The difficulty lies in finding closed-form expressions for
$\alpha_1,\alpha_2$ which will yield a solution $\eta=\eta_{u,1}(z)$ that
satisfies Dirichlet boundary conditions on $0\le z\le 4{\bf K}(m)$.  But
$\eta_{u,1}$, being a ground state, should have no nodes, so it should
extend to a periodic function with period~$8{\bf K}(m)$.  That~is, its
wavenumber should be $\pm\pi/4{\bf K}(m)$.  One of
us~\cite{Maierinpreparation} has found expressions for $\alpha_1,\alpha_2$
in terms of~$\cal E$, which yield a dispersion relation ${\cal E}\mapsto\pm
k$.  By inverting this, we obtain an energy~$\cal E$ corresponding to
$k=\pm\pi/4{\bf K}(m)$, and hence the eigenvalue~$\lambda_{u,1}$.

\begin{figure}[t]
\centerline{\epsfig{file=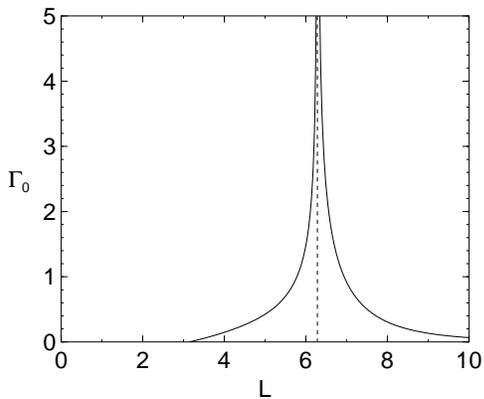,width=2.6in}}
\vskip0.075in
\caption{The Kramers prefactor~$\Gamma_0$ in the Dirichlet case.}
\end{figure}

{\em The Prefactor.}---The Dirichlet-case Kramers prefactor~$\Gamma_0$ can
be computed from the formula~(\ref{eq:prefactor}), using the closed-form
expression~(\ref{eq:quotient}) for the determinant quotient, and the
just-explained technique of calculating the unstable
eigenvalue~$\lambda_{u,1}$.  Figure~3 shows the dependence of~$\Gamma_0$ on
the interval length~$L$.  The divergence at $L=2\pi$ is typical of a
second-order phase transition.

The divergence can be viewed as arising from the bifurcation of the optimal
weak-noise reversal trajectory, rather than the bifurcation of the
transition state on which it terminates.  We~previously found a similar
prefactor divergence in a symmetric two-dimensional nonequilibrium model,
whose optimal trajectory bifurcates but whose transition state does
not~\cite{MaierF}.

A natural question is whether the second-order phase transition is robust.
In~slightly more complicated classical field theories perturbed by
spatiotemporal noise, such transitions may be first-order rather than
second-order, with a discontinuous, rather than diverging, prefactor.
Kuznetsov and Tinyakov~\cite{Kuznetsov97} have studied stationary field
configurations in a sixth-degree Ginzburg--Landau theory, with $\phi +
\alpha \phi^3 - (\alpha+1)\phi^5$ replacing the $\phi - \phi^3$ terms
of~(\ref{eq:Langevin}).  The $\alpha=-1$ case is the theory we have
treated.  If $\alpha>-1$, the periodic instanton branch of the energy
function crosses the sphaleron branch at a nonzero angle.  This would give
rise to a first-order transition.  So, in the $(L,\alpha)$ plane, the
second-order transition point $(2\pi, -1)$ must be the endpoint of a
first-order transition curve.

To sum~up, we have shown that taking spatial extent into account, in simple
models of magnetization reversal induced by weak noise, may yield a rich
structure of activation regimes separated by phase transitions.  The choice
of boundary conditions plays a major role.  We~expect more sophisticated
models, and the phenomenon of field-induced reversal, will display similar
structure.

This research was partially supported by NSF grants PHY-9800979 and
PHY-0099484.

%\bibliographystyle{prsty} \bibliography{general}
%
%\small
%\begin{references}
%\end{references}

\end{document}